\documentclass[aps,superscriptaddress,twocolumn,nofootinbib,12pt,tightenlines]{revtex4}

\makeatletter

\usepackage{natbib}
\usepackage{dcolumn}
\usepackage{graphicx}
\usepackage{amsmath}
\usepackage[hang,small,bf,centerlast]{caption} 
\begin{document}
\title{Market efficiency and the long-memory of supply and demand: Is price impact variable and permanent or fixed and temporary?}


\author{J. Doyne Farmer}
\affiliation{Santa Fe Institute, 1399 Hyde Park Road, Santa Fe, NM 87501}

\author{Austin Gerig}
\affiliation{Department of Physics, University of Illinois at Urbana-Champaign, 1110 West Green Street, Urbana, IL, 61801}
\affiliation{Santa Fe Institute, 1399 Hyde Park Road, Santa Fe, NM 87501}

\author{Fabrizio Lillo}
\affiliation{Dipartimento di Fisica e Tecnologie Relative, viale delle Scienze I-90128, Palermo, Italy}
\affiliation{Santa Fe Institute, 1399 Hyde Park Road, Santa Fe, NM 87501}

\author{Szabolcs Mike}
\affiliation{Santa Fe Institute, 1399 Hyde Park Road, Santa Fe, NM 87501}
\affiliation{Budapest University of Technology and Economics,
H-1111 Budapest, Budafoki \'ut 8, Hungary}

\begin{abstract}
In this comment we discuss the problem of reconciling the linear efficiency of price returns with the long-memory of supply and demand.     We present new evidence that shows that efficiency is maintained by a liquidity imbalance that co-moves with the imbalance of buyer vs. seller initiated transactions.  For example, during a period where there is an excess of buyer initiated transactions, there is also more liquidity for buy orders than sell orders, so that buy orders generate smaller and less frequent price responses than sell orders.  At the moment a buy order is placed the transaction sign imbalance tends to dominate, generating a price impact.  However, the liquidity imbalance rapidly increases with time, so that after a small number of time steps it cancels all the inefficiency caused by the transaction sign imbalance, bounding the price impact.  While the view presented by Bouchaud et al. of a fixed and temporary bare price impact is self-consistent and formally correct, we argue that viewing this in terms of a variable but permanent price impact provides a simpler and more natural view.  This is in the spirit of the original conjecture of Lillo and Farmer, but generalized to allow for finite time lags in the build up of the liquidity imbalance after a transaction.  We discuss the possible strategic motivations that give rise to the liquidity imbalance and offer an alternative hypothesis.  We also present some results that call into question the statistical significance of large swings in expected price impact at long times.
\end{abstract}
\maketitle

\tableofcontents

\section{Introduction}

The fact that supply and demand fluctuations have long-memory, which was independently discovered by Bouchaud et al. (\citeyear{Bouchaud04}) and Lillo and Farmer (\citeyear{Lillo03c}), raises an apparent paradox about compatibility with market efficiency.  The adage that buying drives the price up and selling drives it down is one of the least controversial statements in finance.  The long-memory of supply and demand implies that there are waves of buyer-initiated or seller-initiated transactions that are highly predictable using a simple linear algorithm.  All else being equal, this suggests that price movements should also be highly predictable.  However, from an empirical point of view it is clear that this is not the case -- price movements are essentially uncorrelated.  How can these two facts be reconciled?  

The two original papers on this subject offered two different views to resolve the efficiency paradox.   In an interesting paper Bouchaud et al. (\citeyear{Bouchaud04}) offered the view that one can think in terms of a bare propagator $G_0(t' - t | V, \epsilon)$ that describes the impact of a price change caused by a transaction of volume $V$ and sign $\epsilon$ at time $t$ as it is felt at time $t'$.  The total price impact can be computed by summing up the bare propagators associated with each transaction.  They also assume that the bare impact $G_0$ is {\it fixed}, in the sense that it only depends on the signed volume of the transaction and the time delay, and does not depend on other market properties, such as liquidity.  Given these assumptions, it is clear that the bare propagator has to be temporary, in the sense that it decays to zero as $t' - t \to \infty$.   To see this think about a buy order placed at time $t$, and for simplicity let $\tau = t' - t$. because of the long-memory, $P_b (\tau)/P_s(\tau) \sim \tau^{-\gamma}$, where $0 < \gamma < 1$ is the exponent for the asymptotic decay of the autocorrelation function, $P_b(\tau)$ is the probability of a buyer initiated transaction at time $t + \tau$, and $P_s(\tau)$ is the probability of a seller initiated transaction at time $t + \tau$ (and both are averaged over $t$)\footnote{It is possible to make more accurate predictions by making use of a longer time history, as done in Lillo and Farmer, but to be consistent with Bouchaud et al. we will only consider lags of length one.}. Under their assumptions, to cancel out the inefficiency caused by the long-memory of transactions the sum of all the bare propagators has to go to zero.  Approximating the sum as an integral and taking into account the time decay of the long memory implies that asymptotically $G_0(\tau)$ decays as $G_0 \sim \tau^{(1 - \gamma)/2}$.

An alternate point of view was given by Lillo and Farmer (\citeyear{Lillo03c}), who suggested that this apparent paradox can be explained in terms of permanent price impacts whose size depends on liquidity, where liquidity is defined as the price response to a transaction of a given size.  The liquidity for buying and selling can be different, so that two orders of the same size but opposite sign may generate different average price responses.  This point of view was motivated by related work that suggests that liquidity fluctuations play an essential role in price formation (Farmer et al \citeyear{Farmer04b}, Weber and Rosenow \citeyear{Weber04}, Farmer and Lillo \citeyear{Farmer04}, Lillo and Farmer \citeyear{Lillo05}b, Gillemot, Farmer, and Lillo \citeyear{Gillemot05}).  This body of work demonstrates that the price response to transactions of the same size is highly variable, and that in price formation liquidity fluctuations dominate over fluctuations in transaction size.  Given that liquidity varies with time by many orders of magnitude, it is not surprising that there can be imbalances between the liquidity of buying and selling of less than a factor of two.  As we will see, this is all that is needed to maintain efficiency.

In Section 2.2 of ``Random walks, liquidity molasses, and critical response in financial markets", Bouchaud, Kockelkoren and Potters (\citeyear{Bouchaud04b}), hereafter called BKP, criticize one of the results of Lillo and Farmer (\citeyear{Lillo03c}).   This may be confusing since this result only appeared in the version originally posted on the archive and does not appear at all in the final published paper.  The error in the original version was in the interpretation of an empirical result that suggested that it was possible to remove the excess price impact by the immediate formation of a liquidity imbalance (we will explain what we mean by ``immediate" more precisely in a moment).  Lillo and Farmer thank Marc Potters, who as referee was kind enough to point out the error (and later became a non-anonymous).  The published version of Lillo and Farmer removed this error and added a section showing that immediate liquidity imbalances are large, even if they are not large enough to fully explain the efficiency paradox all by themselves.   The ``liquidity molasses" paper of BKP\footnote{The only caveat is that in Section 2.4 BKP assume that liquidity imbalances are due to price manipulation by market makers; in Section~V we suggest an alternative hypothesis that might cause this.} reflects a significant evolution in point of view from the earlier paper of Bouchaud et al., which emphasized the role of mean-reverting quote changes and did not say much about liquidity imbalance.  In contrast, Figures~7 and 12 of Lillo and Farmer demonstrated that it is liquidity imbalance, rather than mean-reverting price quotes, that is the key to efficiency.  This is echoed in Figure~11 of BKP.  We demonstrate this even more graphically here.

In this comment we summarize some new results that resolve this controversy by explicitly demonstrating how the liquidity imbalance co-varies in time with the long-memory of supply and demand, which we will refer to as the ``transaction imbalance" (defined more precisely later).  Because the liquidity imbalance is initially weaker than the transaction imbalance, there is a initially a nonzero price impact to a transaction.    However, the liquidity imbalance associated with a transaction grows, and after a short time $\tau_c$ becomes large enough to quench any further growth in the price response.  For $\tau > \tau_c$ the liquidity imbalance roughly matches the transaction imbalance, and the price impact remains constant.  Our explanation is thus based on the view that price impacts are permanent but require some time to build after a transaction.  While this is consistent with the formalism introduced by Bouchaud et al (2004), we think that our view is simpler and more natural.  By introducing time dependence, this view contains elements of the original proposals of both Bouchaud et al. and Lillo and Farmer.  

We now explain our new results in more detail.

\section{Time dependence of the liquidity imbalance}


Suppose a transaction of a known sign $\epsilon$ occurs at time $t$, where $\epsilon =  +1$ for a buyer initiated transaction and $\epsilon = -1$ for a seller initiated transaction.  Here and throughout this comment $t$ is measured in transaction time, updated by one unit whenever a transaction occurs.  The size of the expected price impact at time $t + T$  can be written as\footnote{$E[r(T) \epsilon]$ is essentially the same as the average response function $R(l)$ of BKP.  The only difference is that we use log-returns rather than price differences.} 
\begin{eqnarray}
E[r(T)\epsilon] & = & \sum_{\tau=0}^T  \Delta r(\tau) \epsilon \\ \nonumber
& = & \epsilon \sum_{\tau=0}^T [P_+(\tau) R_+ (\tau) + P_- (\tau) R_- (\tau)].
\end{eqnarray}
$\Delta r(\tau) \epsilon$ is the increment by which the size of the expected price impact increases at time $t+\tau$.  Throughout we will assume that all price changes are measured as midprice log returns, i.e. changes in the logarithm of the average of the best quotes for buying and selling.  By summing the log return increments $\Delta r(\tau) \epsilon$ we have defined them to be permanent (though of course positive increments at earlier times might be cancelled by negative increments at later times).  $P_+ (\tau)$ is the expected probability for the transaction at time $t + \tau$ to have sign $\epsilon$ and $R_+$ is the expected log return for transactions at time $t + \tau$ with sign $\epsilon$.  $P_-$ and $R_-$ have similar definitions but with sign $-\epsilon$.

The condition that the price impact increases at time $t + \tau$ is $\Delta r(\tau) \epsilon > 0$, which can alternatively be written
\begin{equation}
\frac{-R_- (\tau)}{R_+ (\tau)} < \frac{P_+ (\tau)}{P_- (\tau)}.
\end{equation}
We will call the term on the right, which reflects the predictability of the transaction signs, the {\it transaction imbalance}.  Similarly, we will call the term on the left, which reflects the asymmetry of the expected price response to buyer vs. seller initiated transactions, the {\it return imbalance}. 

To understand the factors that influence the return imbalance it is useful to decompose the expected return $\Delta r(\tau)$ from directly before the transaction at time $t+\tau$ to directly before the
transaction at time $t+\tau+1$ as\footnote{This was originally done in Equation~(12) of Lillo and Farmer, and is also equivalent to Equation~(12) of BKP.} 
\begin{equation}
\Delta r(\tau) = \Delta r_M (\tau) + \Delta r_Q (\tau).
\end{equation}
$\Delta r_M(\tau)$ is the component of the return that is immediately caused by the receipt of the order that initiates the transaction, while $\Delta r_Q (\tau)$ is everything else, i.e. it includes all changes due to cancellations, or to limit orders that do not cause immediate transactions.  We can similarly decompose $R_+(\tau) = M_+ (\tau) + Q_+ (\tau)$ and $R_- (\tau) = M_- (\tau) + Q_- (\tau)$.   We will call the ratio $-M_- (\tau)/M_+(\tau)$ the {\it liquidity imbalance}.   This makes the term originally introduced by Lillo and Farmer more precise.  As we will demonstrate in the next section, $-R_+(\tau)/R_-(\tau) \approx -M_- (\tau)/M_+(\tau)$ for large values of  $\tau$, so that the return imbalance is fairly well approximated by the liquidity imbalance.

\begin{figure}[ptb]
\begin{center}
\includegraphics[scale=0.4]{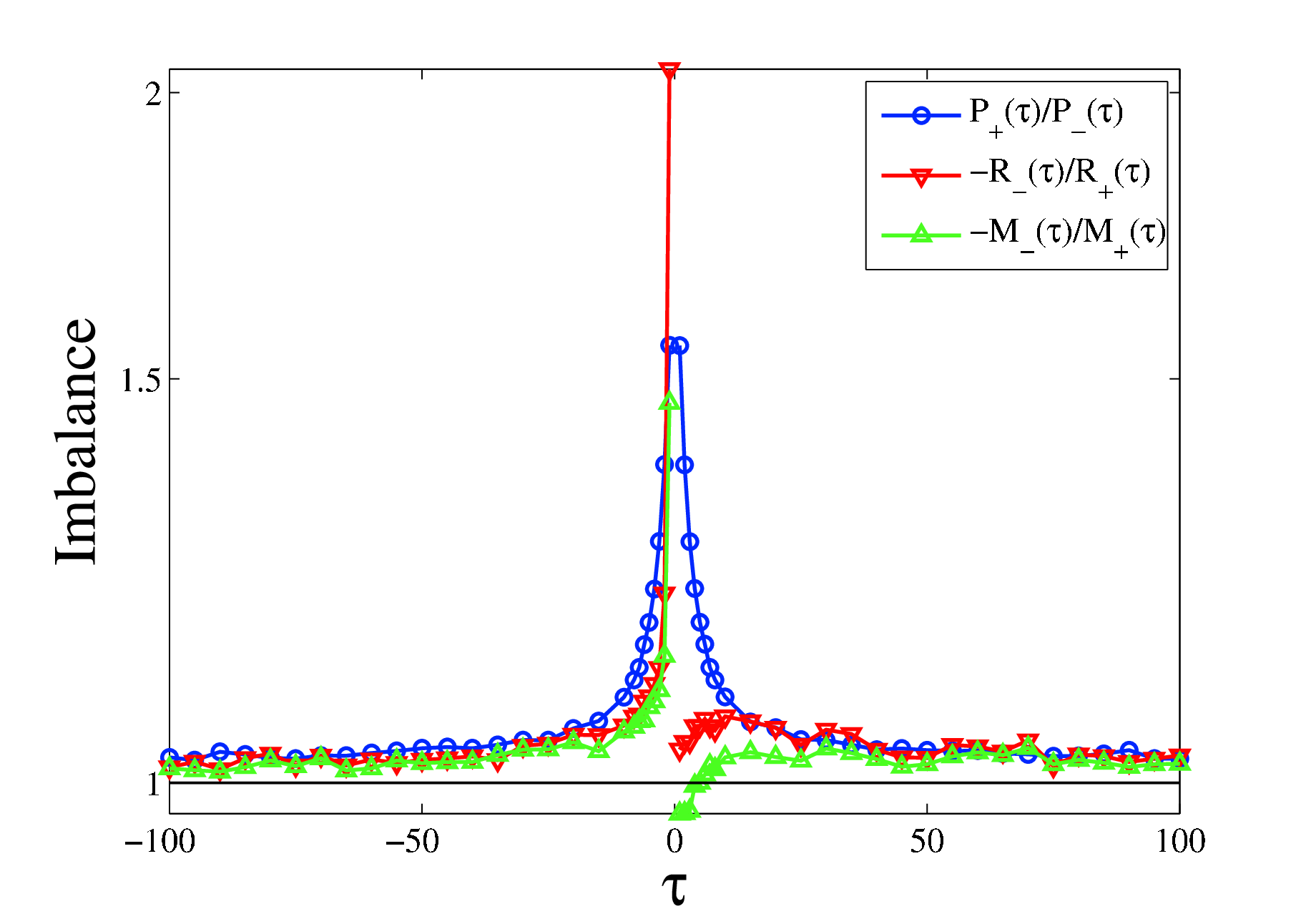}
\end{center}
\caption{A comparison of the transaction imbalance $P_+(\tau)/P_-(\tau)$ (blue circles), return imbalance $-R_-/R_+$ (red down triangles), and liquidity imbalance $-M_- (\tau)/M_+(\tau)$ (green up triangles), for the stock Astrazeneca based on on-book data from $2000-2002$.  Note that we have excluded values at $\tau = 0$, where they diverge or are undefined; the peaks occur at $\tau = -1$.}
\label{liquidityImbalance}
\end{figure}

In Figure~\ref{liquidityImbalance} we compare the transaction imbalance, return imbalance, and liquidity imbalance, using data from the on-book market of the London Stock Exchange for the stock Astrazeneca during the period 2000-2002.  Here and in the following analyses the time $\tau$ is measured in number of transactions, rather than in real time.
The transaction imbalance reaches a peak just before and after the transaction, and is roughly symmetric for positive and negative values of $\tau$. The transaction imbalance is not defined for $\tau=0$ and the slow decay of transaction imbalance is due to the long memory of order sign.  For negative values of $\tau$ the return imbalance rises even more than the transaction imbalance, drops immediately after the transaction, and then builds up again.  The liquidity imbalance behaves similarly but overshoots and becomes negative at small positive values of $\tau$ and then responds more slowly, so that after $\tau_c \approx 40$ it is roughly equal to the transaction imbalance (and the return imbalance).

This result can be interpreted as follows.  Prior to the transaction there is a buildup of the liquidity imbalance, which is amplified by direct movement of quotes. 
Immediately after the transaction the liquidity at the opposite best price is depleted by the transaction, so that the liquidity imbalance dips below one.  This means that the price impact is actually amplified by the liquidity imbalance.  However, the depth of orders at the opposite best price immediately builds, and the liquidity imbalance quickly rises above one.  Even though orders of the same sign continue to be more frequent, their impact is blunted by the buildup of liquidity at the opposite best.  After about $40$ transactions this buildup becomes sufficiently strong so that the imbalance of signs is cancelled by the asymmetry in the price response.

\section{Liquidity imbalance vs. quote reversion}

\begin{figure}[ptb]
\begin{center}
\includegraphics[scale=0.4]{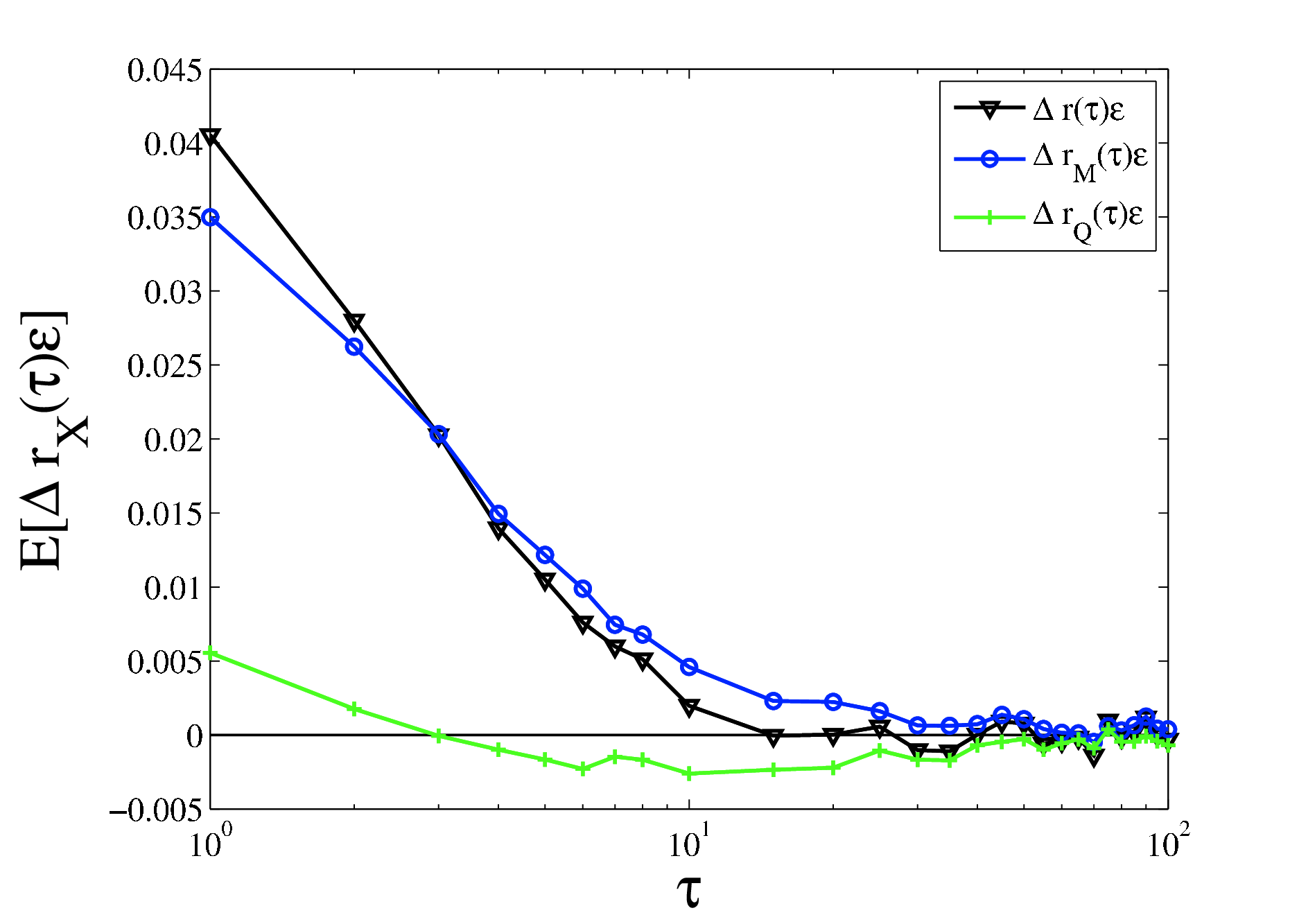}
\end{center}
\caption{A comparison of $\Delta r(\tau)$ (black down triangles), $\Delta
r_M (\tau)$ (blue circles) and $\Delta r_Q (\tau)$ (green crosses) for the
stock Astrazeneca based on on-book data from $2000-2002$.  Returns are in
units of the average spread, which is 0.00154 in logarithmic price.}
\label{impactDynamics}
\end{figure}

To make the time dynamics of the price impact more clear, in
Figure~\ref{impactDynamics} we show $\Delta r(\tau)$, $\Delta r_M (\tau)$
and $\Delta r_Q (\tau)$ as a function of time. 
This shows that price efficiency is mainly the result of the liquidity
imbalance and not mean-reverting quote changes.  $\Delta r(\tau)$ quickly approaches zero due mainly to the
quick decay of $\Delta r_M(\tau)$, which is a result of the liquidity
imbalance.  $\Delta r_Q(\tau)$ begins above zero, which pushes
prices further from efficiency, but eventually does turn negative and helps
to mean-revert the price.  This effect is relatively small, however, and is not consistent across stocks, as we show in a moment for Vodafone - where $\Delta r_Q$ has an overall effect of
making prices less efficient.

\begin{figure}[ptb]
\begin{center}
\includegraphics[scale=0.4]{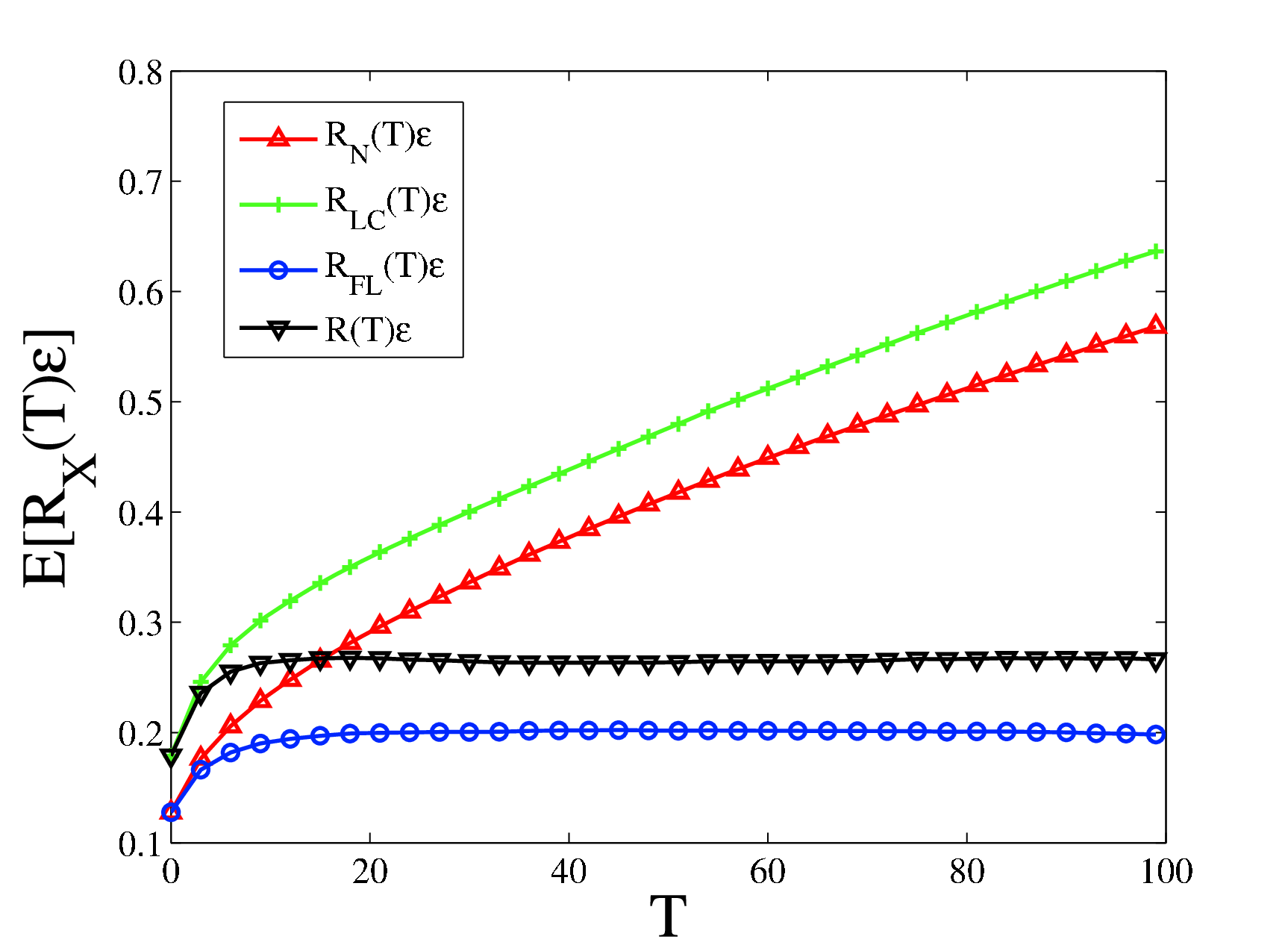}
\end{center}
\caption{A comparison of cumulative price impacts when different effects
are included.  $R_{N}(T)$ (red up triangles) is the naive cumulative price
impact calculated as a sum of fixed price impacts with no liquidity
variation and no variable quote updating.   $R_{LC}(T)$ (green crosses)
includes the effect of non-transaction driven quote changes but does not
include fluctuating liquidity.  $R_{FL}(T)$ (blue circles) includes the
effect of the liquidity imbalance, without including any effect of
non-transaction driven quote changes.  Finally, $R(T)$ (black down
triangles) corresponds to the real cumulative price impact.  All results are
for the stock Vodafone Group based on the on-book data from $2000-2002$.
Returns are in units of the average spread, which is approximately 0.00246 in logarithmic price.}

\label{cumulativeImpact}
\end{figure}

In Figure~\ref{cumulativeImpact} we present cumulative price impacts for Vodafone rather than Astrazeneca.
We show four different curves.  The first of these (red up triangles) shows
the naive cumulative price impact, defined as $R_N(T) = \sum_{\tau = 0}^T
\Delta r_N(\tau) = R_0 \sum_{\tau = 0}^T [P_+(\tau) - P_-(\tau)]$, where
$R_0$ is a constant, the average unconditional absolute midprice impact
measured from immediately before to immediately after a transaction.  This
amounts to assuming that there is no fluctuating liquidity and no
non-transaction driven quote changes -- as each transaction arrives it gives
the price a kick, whose size is independent of time and equal to the average
value.  The second curve (green crosses) adds the quote changing effects of
limit orders and cancellations, $R_{LC}(T) = \sum_{\tau = 0}^T [\Delta
r_N(\tau)+\Delta r_Q(\tau)]$.  This would be observed if all transactions
had impact $R_0$ and non-transaction driven quote changes occurred as usual.
Bouchaud et al originally conjectured that non-transaction driven quote
changes play a major role in mean reverting prices.  We see that in this
case they actually have the opposite effect -- they increase rather
than decrease the cumulative expected return.    The third curve
(blue circles) shows the cumulative price impact if only the effects of
fluctuating liquidity on transaction impacts are included, $R_{FL}(T) =
\sum_{\tau = 0}^T \Delta r_M(\tau)$.  All non-transaction driven quote
changes are excluded.  In this case the return is actually more efficient
than the true cumulative price impact (black down triangles),
$R(T)=\sum_{\tau=0}^T \Delta r(\tau) = \sum_{\tau=0}^T [\Delta r_M(\tau) +
\Delta r_Q (\tau)]$. For Astrazeneca we observe the opposite behavior, i.e. for $\tau$ larger than 20 transactions $R_{LC}$ is smaller than $R_N$ and $R_{FL}$ is larger than the true cumulative impact $R$. However the effect of non-transaction driven quotes is generally very
small, as originally demonstrated by Lillo and Farmer\footnote{See figures 7 and 12.}.

In this section and throughout the comment we have assumed that the imbalance in market order
impacts is due only to the liquidity imbalance, but this is not entirely
true.  An imbalance in transaction volumes can also cause an imbalance in
market order impacts.  In other results that we do not describe here, we
find that the liquidity imbalance is the dominant effect.  We also find that the imbalance in market order impacts is largely
driven by an imbalance in the frequency with which transactions cause
non-zero returns, rather than
asymmetric variations in the size of the returns.  We intend to report these
results in more detail in a future paper (Farmer et al.,
\citeyear{Farmer06}).

\section{Statistical significance of long time behavior}

One of the reasons that Bouchaud et al. argued that the bare impact must be temporary was the observation of mean reversion of price impact at very long times.  In the range $1000 < T < 5000$, they sometimes observe that price impact becomes close to zero or even becomes negative, while in other cases it increases dramatically.  They argue that this reflects fluctuations in the balance between liquidity taking and liquidity providing -- in some stocks and in some periods the liquidity providers are stronger, and in others the liquidity demanders are stronger.  

A key issue that is not properly addressed is statistical significance.  Are these deviations real, or are they just statistical fluctuations, causing random variations from stock to stock?  In Figure~10 of Bouchaud et al. there are error bars on the price impact, which appear to indicate that the reversion of the impact that is empirically observed is statistically significant.    However, these are based on standard errors.  This can be problematic when long-memory is a possibility, in which case under the assumption of a normal distribution the one standard deviation errors are of size
\begin{equation}
E \approx \frac{\sigma}{n^{(1-H)}},
\label{significanceFormula}
\end{equation}
where $n$ is the number of non-overlapping observations, $H = 1 - \gamma/2$ is the Hurst exponent, and $\sigma$ is the standard deviation of the random process (Beran, \citeyear{Beran94}).  When $H = 1/2$ this reduces to the expression for standard error, but when $H > 1/2$ there is long-memory and errors can be much larger than one might naively expect.  Volatility is a long-memory process with Hurst exponents in the neighborhood of $H \approx 0.75$ (Gillemot, Farmer, and Lillo, \citeyear{Gillemot05}).  Thus, errors decrease roughly as the fourth root rather than the square root of the number of non-overlapping observations.

This problem is compounded by the fact that the increments of the response function are overlapping.  Thus, in a data series with a million points, at lags of $T = 1000$ there are only 1000 non-overlapping intervals.   To avoid crossing daily boundaries Bouchaud et al and BKP only used events within the same day, which decreases the number of independent intervals for large values of $T$ even more.  Because of the overlapping intervals and long-memory effects, errors at nearby times are highly correlated -- once the price impact function becomes large (or small), it is likely to remain so for some time simply because of the correlated errors.

\begin{figure}[ptb]
\begin{center}
\includegraphics[scale=0.4]{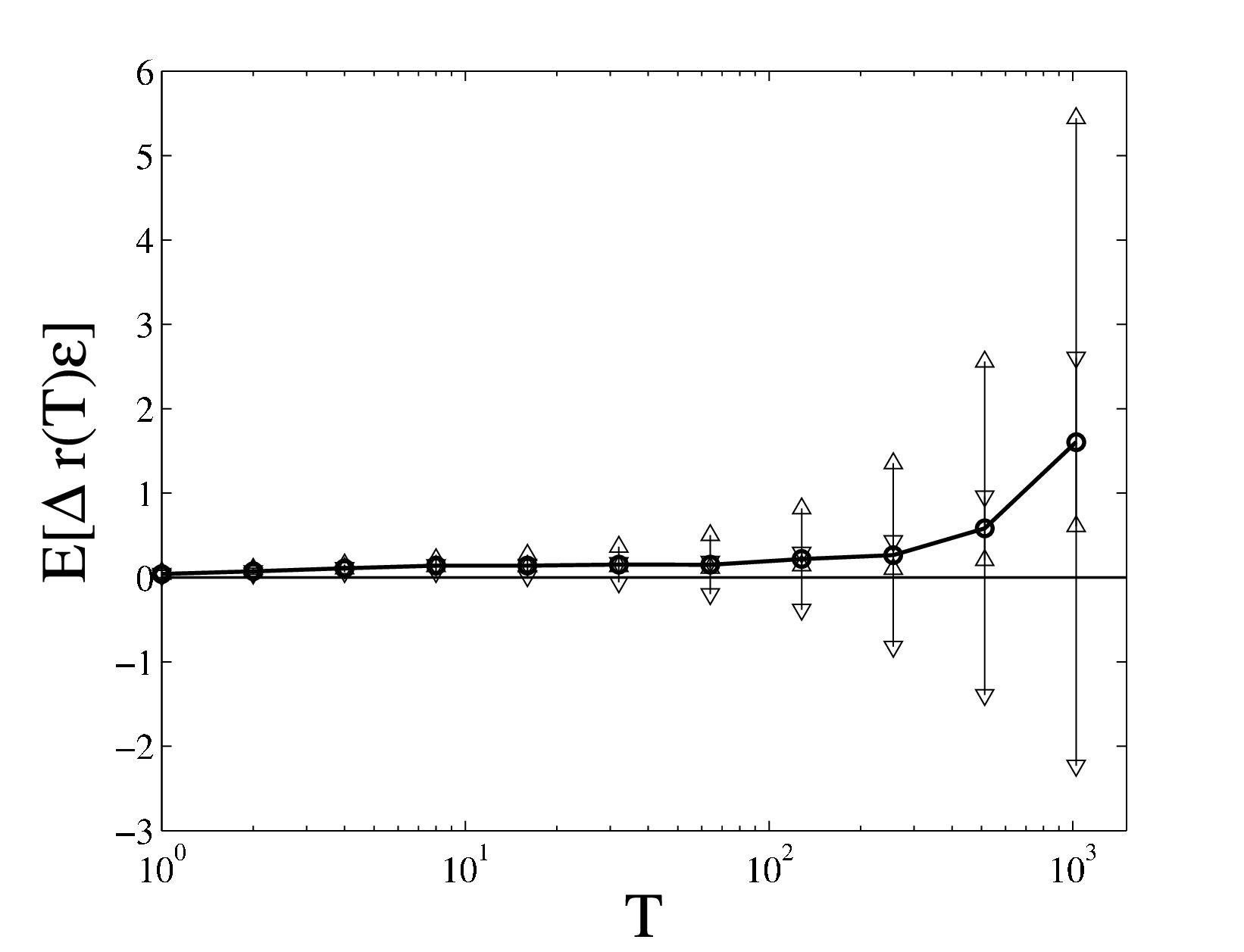}
\end{center}
\caption{A crude estimate of the statistical errors in the price impact at different times for Astrazeneca based on the on-book data for 2000-2002 based on equation~\ref{significanceFormula} (triangle up).  The circles are the estimated mean impact.  For comparison the figure shows also standard errors (triangle down). Returns are shown in units of the average spread.}
\label{statisticalSignificance}
\end{figure}

Properly resolving the question of whether or not the large excursions of the price impact at long times are statistically significant is a difficult job that is beyond the scope of this comment.  But just to illustrate the problem, in Figure~\ref{statisticalSignificance} we make a crude estimate of error bars for the price impact.
We do this by measuring the standard deviation $\sigma(T)$ of impacts at several different time intervals $T$ and estimating the one standard deviation statistical error using Equation~\ref{significanceFormula}.    The number of non-overlapping intervals is $n = N/T$, where $N$ is the total number of points in the data set.   Because the standard deviation of the impact grows roughly as $\sqrt{T}$, the error bars therefore grow roughly as $T^{3/2}$.  As a result, at $T \approx 100$ the error bars become bigger than the estimated values.  The figure also shows the error bars computed assuming standard errors, i.e. $H=1/2$. In this case the error bars are smaller than the mean values. This procedure is crude -- it assumes normality, which is a poor assumption for price impacts, and the use of overlapping intervals should give a somewhat more accurate estimate than this procedure would indicate.  But at least it takes long-memory into account, and it indicates that a more careful study is needed to determine whether the divergences in the price impact observed at long times by Bouchaud et al and BKP are statistical fluctuations.

\section{Strategic interpretation}

While the discussion in Sections~2 and 3 gives insight into {\it how} efficiency is maintained, it does not address the strategic motivations for {\it why} it is maintained.  Is the motive profit seeking, liquidity demanding, risk reduction, or is there some other cause?  If it is profit seeking, then what kind of strategy causes the inefficiency to be removed?  For example, suppose there is a buying wave.  We have demonstrated that the main force stopping a run up in prices is the rapid formation of a liquidity imbalance.  Is this imbalance created intentionally by market makers, or is it a side effect of some other behavior?

Bouchaud et al and BKP speculate that this is due to intentional controlling of prices by market makers.  We fully agree that long-memory is bad news for market makers, but whether they can solve it by controlling the price is not so obvious.   The main reasons for our skepticism are inventory control and the need for cooperative behavior if market making is competitive.  When a buying wave occurs, to keep the price from rising by maintaining a liquidity imbalance the market makers must sell and thereby absorb most of the buying wave themselves.  This causes them to accumulate a negative inventory.  To flatten this inventory they eventually have to buy, which will tend to drive the price up.  Alternatively, they might wait for the next selling wave, but this could be a long time coming -- once in a buying wave, because of the long-memory, the most likely future is more buying waves.  Also, it might not match the previous buying wave in size, still leaving them with a net negative inventory.  In a competitive market making situation (which exists for both the London and Paris Stock Exchanges), inventory constraints gives strong incentives for market makers to try to free ride on each other.  Once they have fulfilled their inventory goals, free riders will back their quotes off to let someone else absorb the rest of the dangerous negative inventory.  It seems that the Nash equilibrium would be for everyone to defect.  Taken together, these issues could make it difficult for market makers to control the price, particularly in a competitive situation. 

An alternate hypothesis is that liquidity providing and liquidity taking are self-reinforcing.  Liquidity is not only provided by market makers -- it is also provided by directional traders who are simply more patient than their liquidity taking cousins.  By directional trader we mean someone who at any given time either wants to buy or wants to sell but never wants to do both.  Because they are more patient they use limit orders, but to avoid broadcasting their true intentions they only place orders of a size that they think the market can absorb, and place new orders only after their existing orders are hit.  When the patient sellers observe an impatient buying wave of liquidity takers, they increase the size of their sell orders and replace them more frequently.  The impatient buyers see the resulting increase in liquidity, which stimulates them to submit more market orders.  This gives rise to bi-directional causality -- buy liquidity taking causes sell liquidity providing, and  sell liquidity providing causes buy liquidity taking (and similarly with buying and selling reversed).  This scenario is indirectly suggested by the lack of a clear lead-lag relationship between liquidity taking and liquidity providing in Figure~1.  The key difference is that under this scenario the inventory swings are absorbed by traders who want to absorb them, rather than by market makers who may not want to absorb them if they are too large.  Market makers will participate in liquidity provision to the extent that it conforms to their inventory constraints, but hide behind directional liquidity providers when it does not. 

Such bursts of buying and selling may be triggered by the occurrence of large hidden orders, according to the theory developed by Lillo, Mike and Farmer (\citeyear{Lillo05b}).   When traders receive large hidden orders, some may choose to execute them via a sequence of smaller limit orders and some via a sequence of smaller market orders.  In either case, under this scenario such a hidden order will trigger a response of other hidden orders of the opposite sign and opposite liquidity type, amplifying the exogenous input of large hidden orders.

\section{Concluding remarks}

We have demonstrated that efficiency in the face of long-memory in transaction signs is maintained by the creation of a liquidity imbalance that asymmetrically alters the expected return triggered by a transaction.  This liquidity imbalance takes a fairly short time to build up -- for the cases we have observed here most of the build up happens in about 5 transactions (on average a few minutes for Astrazeneca), and the build up is completed by about 40 transactions (on average about 25 minutes for Astrazeneca).  As a result the price impact rises to roughly $0.25 - 0.3$ of the average spread and then levels out.  Under this view, price impact is variable and permanent.  When a transaction happens, it affects the next 40 or so transactions by building up their liquidity, until the liquidity imbalance matches the transaction imbalance, and after that the two move in tandem.  This generalizes the proposal originally made by Lillo and Farmer by allowing some time (but not much) for the growth of the liquidity imbalance.

The fixed bare propagator approach of Bouchaud et al does not formally contradict this view -- in fact in BKP they more or less embrace it.  However, to us this interpretation does not seem to arise naturally from their formalism.  By proposing a bare propagator that is fixed, their phenomenological theory requires that it be temporary in order to eventually blunt the growth of the price impact. This is formally correct -- once the liquidity imbalance is set up, for $\tau > \tau_c$ the time decay of the bare propagator is needed to match the time decay of the transaction imbalance.  We think that the results we have presented here shows that it is simpler and more natural to think in terms of a permanent but variable impact function.  This is perhaps just a matter of taste -- one can think about this phenomenon in either way.  In any case, in this comment we have presented some new results that make the explicit mechanisms that enforce market efficiency clearer.

\acknowledgments We would like to thank Barclay's Bank for supporting this research.  F.L acknowledges partial support from MIUR research project `` Dinamica di altissima frequenza nei mercati finanziari", MIUR-FIRB research project RBNE01CW3M and NEST-DYSONET 12911 EU project. We would also like to thank J-P. Bouchaud and Marc Potters for useful discussions.
 
\bibliographystyle{plainnat}

\end{document}